\input pictex

\parindent0pt
\baselineskip=1.5\baselineskip

\footline{\hss{\sl \ Page \folio}}

\def\hor{\widehat{OR}}
\def\tor{\widetilde{OR}}
\def\frac#1#2{{#1\over#2}}
\def\binom#1#2{{#1\choose#2}}
\def\rkonfl#1#2#3#4{{}_2{\widetilde F}_1\left[#1,#2;#3;#4\right]}


\centerline{\bf A New Confidence Interval for the Odds Ratio}

\vskip1truecm
\centerline{\bf Zofia Zieli{\'n}ska-Kolasi{\'n}ska}
\medskip
\centerline{Department of Mathematics and Physics, Siedlce University of Natural Sciences and Humanities}
\centerline{Stanislawa Konarskiego 2, 08-110 Siedlce, Poland}
\medskip
\centerline{e-mail: zofia.zielinska-kolasinska@uph.edu.pl}
\bigskip
\centerline{\bf Wojciech Zieli{\'n}ski}
\medskip
\centerline{Department of Econometrics and Statistics, Warsaw University of Life Sciences}
\centerline{Nowoursynowska 159, 02-776 Warszawa, Poland}
\centerline{and}
\centerline{Department of the Prevention of Environmental Hazards and Allergology, Medical University of Warsaw}
\centerline{Banacha 1a, 02-097 Warszawa, Poland}
\medskip
\centerline{e-mail: wojciech\_zielinski@sggw.edu.pl}
\centerline{http://wojtek.zielinski.statystyka.info}
\vskip1truecm

\bigskip
{\bf Abstract} We consider the problem of interval estimation of the odds ratio. An asymptotic confidence interval is widely applied in practise: economic, medical, sociological etc. research. Unfortunately that confidence interval has a poor coverage probability: it is significantly smaller than the nominal confidence level. In this paper a new confidence interval is proposed. For the construction only information on sample sizes and sample odds ratio is sufficient. The coverage probability of the proposed confidence interval is at least the nominal confidence level.
\bigskip

{\bf Keywords} confidence interval, odds ratio
\bigskip

{\bf AMS Subject Classification} Primary 62F25; Secondary 62P10, 62P20

\bigskip
{\bf 1. Introduction}
\bigskip

In many practical sciences such as economy, medicine, sociology etc. dichotomous variate is observed. Such variate is to be compared in two independent groups. Commonly used is difference of two fractions (the risk difference), the ratio of two proportions (the relative risk) and the odds ratio. The relative risk and the odds ratio are relative measurements for comparison of two variates, while  the risk difference is an absolute measurement.

The odds ratio is one of the parameters commonly used in such comparisons, especially in two-arm binomial experiments. This indicator was firstly applied by Cornfield (1951). The literature devoted to the analysis of odds ratio and its estimators is very rich, see e.g. Encyclopedia of Statistical Sciences (2006) Volume 9, pp. 5722-5726 and the literature therein.

However, the problem is in the interval estimation. In general there are two approaches to the problem. The first one consists of the analysis of $2\times2$ tables (Edwards 1963, Gart 1971, Thomas 1971). The second approach is based on logistic model in which the odds ratio has a direct relationship with the regression coefficient (Gart 1971, McCullagh 1980, Morris \& Gardner 1988). Wang \& Shan (2015) constructed exact confidence interval for the odds ratio based on another approach. Namely they applied the so called rank function. Very exhaustive review of different confidence intervals for the odds ratio may be found in Andr{\'e}s et.al (2020). Unfortunately all those confidence intervals have one very important disadvantage: their real probability of coverage is significantly smaller than the nominal one. It is in contradiction to Neyman (1934, p. 562) definition of a confidence interval. Hence the risk of a wrong conclusion (i.e. overestimation or underestimation) is greater than the assumed one and unluckily remains unknown.

The most commonly used in applications is an asymptotic interval for odds ratio derived from logistic model (formula $(S)$ in Section 3). This asymptotic interval is widely used in different statistical packages. There are also many internet scripts for calculating the asymptotic confidence interval (see e.g.  http://www.hutchon.net/ConfidOR.htm). Unfortunately this confidence interval has some statistical disadvantages discussed in Section 3. To avoid those disadvantages a new confidence interval is proposed. The idea of construction is similar to the idea of construction of the confidence interval for the difference of two probabilities of success (the risk difference) proposed by Zieli{\'n}ski (2020).

In Section 2 a new confidence interval is constructed. In Section 3 some disadvantages of the asymptotic confidence interval are discussed. Final conclusions are given in Section 4.

\bigskip
{\bf 2. A new confidence interval}
\bigskip

Consider two independent r.v.'s $\xi_A$ and $\xi_B$ distributed as $Bin(n_A,p_A)$ and $Bin(n_B,p_B)$, respectively. The problem is in estimating the odds ratio:
$$OR=\frac{(p_A/(1-p_A))}{(p_B/(1-p_B))}=\frac{p_A}{(1-p_A)}\cdot\frac{(1-p_B)}{p_B}.$$
Let $n_{A1}$ and $n_{B1}$ be observed numbers of successes. The data are usually organized in a $2\times2$ table:
$$\vbox{\tabskip1em\offinterlineskip\halign{
\strut\hfil#\hfil&#\vrule width1pt&\hfil$#$\hfil&#\vrule&\hfil$#$\hfil&#\vrule width1pt&\hfil$#$\hfil\cr
&&\hbox{success}&&\hbox{failure}&& \cr\noalign{\hrule height1pt}
Group $A$&&n_{A1}&&n_{A0}&&n_A\cr
Group $B$&&n_{B1}&&n_{B0}&&n_B\cr\noalign{\hrule height1pt}
&&n_1&&n_0&&n\cr
}}$$

The standard estimator of $OR$ is as follows:
$$\tor=\frac{n_{A1}}{n_A-n_{A1}}\cdot\frac{n_B-n_{B1}}{n_{B1}}\eqno{(\star)}$$
\bigskip

Usually, the problem of estimating an odds ratio is considered in the following statistical model:
$$\left(\left\{0,1,\ldots,n_A\right\}\times\left\{0,1,\ldots,n_B\right\},\left\{Bin(n_A,p_A)\cdot Bin(n_B,p_B), (p_A,p_B)\in(0,1)\times(0,1)\right\}\right).$$
Since we are interested in estimating the odds ratio $OR$, consider now a new statistical model. This model is the one-parameter model: the odds ratio is an unknown parameter
$$\left({\cal X},\left\{F_r, 0\leq r\leq+\infty\right\}\right),$$
where
$${\cal X}=\left\{\frac{n_{A1}}{n_A-n_{A1}}\cdot\frac{n_B-n_{B1}}{n_{B1}}: n_{A1}\in\{0,1,\ldots,n_A\},n_{B1}\in\{0,1,\ldots,n_B\}\right\}.$$
The cumulative distribution functions (CDF) $F_r(\cdot)$ are defined as follows.

Note that the estimator $\tor$ given by formula $(\star)$ is undefined for $n_{A1}=n_A$ or $n_{B1}=0$. We extend the definition of the estimator of the odds ratio in the following way:
$$\hor=\cases{
0,&for $(n_{A1}=0,n_{B1}\geq1)$ or $(n_{A1}\leq n_A-1,n_{B1}=n_B)$\cr
+\infty,&for $(n_{A1}=n_A,1\leq n_{B1}\leq n_B-1)$ or $(n_{A1}\geq1,n_{B1}=0)$\cr
1,&for $(n_{A1}=0,n_{B1}=0)$ or $(n_{A1}= n_A,n_{B1}=n_B)$\cr
\hbox{formula }(\star),&elsewhere\cr}
\eqno{(\star\star)}$$


To find the distribution of $\hor$ note that for a given odds ratio equal to $r>0$
$$p_B=\frac{p_A}{p_A+r(1-p_A)};\ 1-p_B=\frac{r(1-p_A)}{p_A+r(1-p_A)}.$$
The probability of observing $\xi_A=n_{A1}$ and $\xi_B=n_{B1}$ equals
$$P_{p_A,p_B}\left\{n_{A1},n_{B1}\right\}=\binom{n_A}{n_{A1}}p_A^{n_{A1}}(1-p_A)^{n_A-n_{A1}}\binom{n_B}{n_{B1}}p_B^{n_{B1}}(1-p_B)^{n_B-n_{B1}}.$$
Equivalently
$$P_{r,p_A}\left\{n_{A1},n_{B1}\right\}=r^{n_B-n_{B1}}\binom{n_A}{n_{A1}}\binom{n_B}{n_{B1}}\frac{p_A^{n_{A1}+n_{B1}}(1-p_A)^{n_A+n_B-n_{A1}-n_{B1}}}{(p_A+r(1-p_A))^{n_B}}.$$

The probability $p_A$ is a nuisance parameter and will be eliminated by an appropriate integration
$$\eqalign{
P_{r}\left\{n_{A1},n_{B1}\right\}&=\int_0^1 P_{r,p_A}\left\{n_{A1},n_{B1}\right\}dp_A\cr
&=(n_{A}+n_{B})!\frac{\binom{n_A}{n_{A1}}\binom{n_B}{n_{B1}}}{\binom{n_A+n_B}{n_{A1}+n_{B1}}}\left(\frac{1}{r}\right)^{n_{B1}}\rkonfl{n_B}{n_{A1}+n_{B1}+1}{n_A+n_B+2}{1-\frac{1}{r}},\cr}$$
where
$$\rkonfl{x}{y}{z}{t}=\frac{1}{\Gamma(z-y)\Gamma(y)}\int_0^1u^{y-1}(1-u)^{z-y-1}(1-ut)^{-x}du\ (\hbox{for }z>y>0)$$
is the regularized confluent hypergeometric function.
The CDF of $\hor$ equals (for $t\geq0$)
$$F_r(t)=P_r\left\{\hor\leq t\right\}=\sum_{n_{A1}=0}^{n_A}\sum_{n_{B1}=0}^{n_B}P_{r}\left\{n_{A1},n_{B1}\right\}{\bf1}\left(\hor\left(n_{A1},n_{B1}\right)\leq t\right),$$
where ${\bf1}\left(q\right)=1$ when $q$ is true and $=0$ elsewhere.

Since $\hor$ is given by formula $(\star\star)$ the CDF may be written as
$$\eqalign{
F_r(t)&=\sum_{n_{A1}=0}^{n_A-1}\sum_{n_{B1}=h(n_{A1})}^{n_B}P_{r}\left\{n_{A1},n_{B1}\right\}\cr
&=(n_A + n_B)!\sum_{n_{A1}=0}^{n_{A}-1}\sum_{n_{B1}=h(n_{A1})}^{n_B}\frac{\binom{n_A}{n_{A1}}\binom{n_B}{n_{B1}}}{\binom{n_A+n_B}{n_{A1} + n_{B1}}}\left(\frac{1}{r}\right)^{n_{B1}}\rkonfl{n_B}{n_{A1} + n_{B1} + 1}{n_A + n_B + 2}{1 -\frac{1}{r}}\cr
}$$
where $$h(n_{A1})=\cases{\left\lceil\frac{n_B}{t\left(\frac{n_A}{n_{A1}}-1\right)+1}\right\rceil,&for $n_{A1}\geq1$,\cr 0,&for $n_{A1}=0$,\cr}$$
(here $\lceil x\rceil$ denotes the smallest integer not less than $x$).


The family $\left\{F_r,r\geq0\right\}$ is stochastically ordered, i.e. for a given $t>0$
$$F_{r_1}(t)\geq F_{r_2}(t)\hbox{ for } r_1\leq r_2.$$
It follows from the fact that for a given $n_{A1}$, $n_{B1}$ and $p_A$ the probability $P_{r,p_A}\left\{n_{A1},n_{B1}\right\}$ is the decreasing function of odds ratio $r$ and hence $P_{r}\left\{n_{A1},n_{B1}\right\}$ is also decreasing in $r$.

By $G_r(t)$ denote the probability $P_r\left\{\hor< t\right\}$.
Let $\gamma$ be the given confidence level and let $\hat r$ be the observed odds ratio. The confidence interval for $r$ takes on the form
$$\left(Left\left(\hat r\right),\ Right\left(\hat r\right)\right),\eqno{(M)}$$
where
$$Left\left(\hat r\right)=\cases
{0,& ${\hat r}=0$,\cr
0,& if $\lim_{r\to0}G_r\left(\hat r\right)<(1+\gamma)/2$,\cr
r_*,& $r_*=\max\left\{r: G_{r}\left(\hat r\right)\geq(1+\gamma)/2\right\}$,\cr
}$$
and
$$Right\left(\hat r\right)=\cases
{\infty,& ${\hat r}=\infty$,\cr
\infty,& if $\lim_{r\to\infty}F_r\left(\hat r\right)>(1-\gamma)/2$,\cr
r^*,&$r^*=\min\left\{r: F_{r}\left(\hat r\right)\leq(1-\gamma)/2\right\}$.\cr
}$$

\bigskip

{\bf Theorem.} For $n_A>\frac{2}{1-\gamma}-1$ the confidence interval for the odds ratio is two-sided and is one-sided otherwise.

For the proof see Appendix 1.

\bigskip

If $\hat r$ is the observed odds ratio then the confidence interval for $r$ takes on the following form:
$$\eqalign{
\hbox{for }\hat r\in[0,1):&\cases{\langle0,r^*),& for $n_A\leq\frac{2}{1-\gamma}-1$,\cr (r_*,r^*),& for $n_A>\frac{2}{1-\gamma}-1$,\cr}\cr
\hbox{for }\hat r\in[1,+\infty):&\cases{(r_*,+\infty),& for $n_A\leq\frac{2}{1-\gamma}-1$,\cr (r_*,r^*),& for $n_A>\frac{2}{1-\gamma}-1$,\cr}\cr
}$$
where $r_*$ and $r^*$ are given by formula $(M)$.

\bigskip
Minimal sample sizes $n_A$ for which two-sided confidence interval exists are given in Table 1.
$$\vbox{\tabskip1em\offinterlineskip\halign{
\strut\hfil$#$\hfil&#\vrule&&\hfil$#$\hfil\cr
\multispan{6}{\bf Table 1.} Minimal sample size\hfill\cr\noalign{\vskip5pt}
\gamma&&0.9&0.95&0.99&0.999\cr\noalign{\hrule}
n_A&&20&40&200&2000\cr
}}$$

\bigskip

For a given $r>0$ the coverage probability, by construction, equals at least $\gamma$. In Figure 1 there is shown the coverage probability for $n_A=60$, $n_B=70$ and $\gamma=0.95$. On the $x$-axis the value $r$ of the odds ratio is given and on the $y$-axis the probability of coverage is shown. The coverage probabilities are calculated, not simulated.

$$\beginpicture
\setcoordinatesystem units <30truemm,58truecm>
\setplotarea x from 0 to 2, y from 0.94 to 1.00
\axis bottom
 ticks numbered from 0 to 2 by 0.50 /
\axis left
 ticks numbered from 0.94 to 1.00 by 0.01 /
\plot "Pokrycie_moje.tex"
\setdashpattern<1.5truemm,1.5truemm>
\plot 0 0.95 2 0.95 /
\put {{\bf Figure 1.} Coverage probability of $(M)$.} at 1.00 0.925
\endpicture
$$

\bigskip

\bigskip
{\bf Remark.} The above considerations are made for $A$ versus $B$. It is obvious that
$$OR(A\hbox{ vs }B)=\frac{1}{OR(B\hbox{ vs }A)}.$$
It is easily seen that the new confidence interval has the following natural property:
$$Left(A\hbox{ vs }B)=\frac{1}{Right(B\hbox{ vs }A)}\quad\hbox{and}\quad Right(A\hbox{ vs }B)=\frac{1}{Left(B\hbox{ vs }A)}.$$
In case of considering $B$ versus $A$ in the Theorem the sample size $n_A$ should be changed to $n_B$.

\bigskip
{\bf 3. Standard confidence interval}
\bigskip

Estimating the odds ratio is one of the crucial problems in medicine, biometrics etc. The most widely used confidence interval at the confidence level $\gamma$ is of the form
$$\left(\tor\cdot\exp\left(u_{\frac{1-\gamma}{2}}\sqrt{\frac{1}{n_{A1}}+\frac{1}{n_{A0}}+\frac{1}{n_{B1}}+\frac{1}{n_{B0}}}\right),
\tor\cdot\exp\left(u_{\frac{1+\gamma}{2}}\sqrt{\frac{1}{n_{A1}}+\frac{1}{n_{A0}}+\frac{1}{n_{B1}}+\frac{1}{n_{B0}}}\right)\right),\eqno{(S)}$$
where $u_{\delta}$ denotes the $\delta$ quantile of $N(0,1)$ distribution. In the above formula the estimator $\tor$ is given by $(\star)$.
Unfortunately this confidence interval has at least three disadvantages. They are as follows.

\bigskip
{\bf 1.} Confidence interval $(S)$  does not exist if at least one of $n_{A0}$, $n_{A1}$, $n_{B0}$ or $n_{B1}$ equals zero.

\bigskip
{\bf 2.} The coverage probability of c.i. $(S)$ is less than the nominal one. In Figure 2 the coverage probability is shown for $n_A=60$, $n_B=70$ and $\gamma=0.95$ (the value $r$ of odds ratio is given on the $x$-axis and  the coverage probability is given on the $y$-axis).
$$\beginpicture
\setcoordinatesystem units <30truemm,20truecm>
\setplotarea x from 0 to 2, y from 0.80 to 1.00
\axis bottom
 ticks numbered from 0 to 2 by 0.50 /
\axis left
 ticks numbered from 0.80 to 1.00 by 0.05 /
\plot "Pokrycie_std.tex"
\setdashpattern<1.5truemm,1.5truemm>
\plot 0 0.95 2 0.95 /
\put {{\bf Figure 2.} Coverage probability of $(S)$.} at 1.00 0.76
\endpicture
$$
The probability of wrong conclusion, i.e. of overestimation or underestimation is greater than the assumed $0.05$. It means that the true value of odds ratio may be smaller than the left end of the confidence interval $(S)$ or greater than its right end. The risk of such event is greater than the nominal $0.05$ and unfortunately remains unknown. Note that this is in contradiction to Neyman (1934, p. 562) definition of a confidence interval.

\bigskip
{\bf 3.} The standard asymptotic confidence interval requires the knowledge of sample sizes as well as sample proportions in each sample. Unfortunately it may lead to misunderstandings. Namely, suppose that six experiments were conducted. In each experiment two samples of sizes sixty and seventy respectively, were drawn ($n_1=60$, $n_2=70$). The resulting numbers of successes are shown in Table 2 (the first two columns).
\midinsert
$$\vbox{\tabskip1em minus0.9em\offinterlineskip\halign to0.5\hsize{
\strut\hfil$#$\hfil&&#\vrule&\hfil$#$\hfil\cr
\multispan{9}\strut{\bf Table 2.} Confidence intervals in six experiments\hfill\cr\noalign{\vskip5pt}
n_{A1}&&n_{B1}&&\tor&&left&&right\cr\noalign{\hrule}
6&&14&&0.4444&&0.1592&&1.2410\cr
8&&18&&0.4444&&0.1776&&1.1122\cr
15&&30&&0.4444&&0.2095&&0.9428\cr
24&&42&&0.4444&&0.2199&&0.8985\cr
36&&54&&0.4444&&0.2078&&0.9506\cr
48&&63&&0.4444&&0.1627&&1.2141\cr
}}$$
\endinsert
It is seen that the sample odds ratio (the third column) is the same in all experiments, but the confidence intervals are quite different. Moreover, for example in the first experiment it may be claimed that the population odds in groups $A$ and $B$ may be treated as equal, while in the fourth one such a conclusion should not be drawn.

\bigskip
{\bf 4. Conclusions}
\bigskip

In this paper a new confidence interval for the odds ratio is proposed. The confidence interval is based on the exact distribution of the sample odds ratio, hence it works for large as well as for small samples. The coverage probability of that confidence interval is at least the nominal confidence level, in contrast to the asymptotic confidence intervals known in the literature. It must be noted that the information on the sample sizes and the sample odds ratio is sufficient for constructing the new confidence interval. Unfortunately, no closed formulae for the ends of the confidence interval are available. However, for given $n_A$, $n_B$ and observed $\hor$ the ends may be easily numerically computed with the aid of the standard software such as R, Mathematica etc (see Appendix 2).

Since the proposed confidence interval may be applied for small as well as for large sample sizes, it may be recommended for practical use.


\bigskip
{\bf References}
\bigskip

\begingroup
\parskip6pt
\baselineskip=12pt

Andr{\'e}s, A. M., Tejedor, I. H., Hern{\'a}ndez, M. A. (2020). Two-tailed asymptotic inferences for the odds ratio in prospective and retrospective studies: evaluation of methods of inference, Journal of Statistical Computation and Simulation, 90: 138-156, DOI: 10.1080/00949655.2019.1673751

Cornfied, J. (1951). A Method of Estimating Comparative Rates from Clinical Data. Applications to Cancer of the Lung, Breast, and Cervix. JNCI: Journal of the National Cancer Institute. 11: 1269-1275, DOI: 10.1093/jnci/11.6.1269

Edwards, A. W. F. (1963). The Measure of Association in a $2\times2$ Table. Journal of the Royal Statistical Society. Ser. A. 126: 109-114. DOI: 10.2307/2982448.

Encyclopedia of Statistical Sciences (2006), Wiley \& Sons

Gart, J. J. (1971). The comparison of proportions: a review of significance tests, confidence intervals, and adjustments for stratification. Review of the International Statistical Institute. 39: 148-169.

Lawson, R. (2004). Small Sample Confidence Intervals for the Odds Ratio. Communications in Statistics - Simulation and Computation. 33: 1095-1113, DOI: 10.1081/SAC-200040691.

Morris, J. A., Gardner M. J. (1988). Calculating confidence intervals for relative risks (odds ratios) and standardised ratios and rates. British Medical Journal. 296: 1313-6. DOI: 10.1136/bmj.296.6632.1313.

McCullagh, P. (1980). Regression Models for Ordinal Data. Journal of the Royal Statistical Society. Ser. B. 42: 109-142.

Neyman, J. (1934). On the Two Different Aspects of the Representative Method: The Method of Stratified Sampling and the Method of Purposive Selection, Journal of the Royal Statistical Society. 97: 558-625.

Thomas, D. G. (1971). Algorithm AS-36: exact confidence limits for the odds ratio in a $2\times2$ table. Applied Statistics. 20: 105-110.

Wang, W., Shan G. (2015). Exact Confidence Intervals for the Relative Risk and the Odds Ratio. Biometrics. 71: 985-995 DOI: 10.1111/biom.12360.

Zieli{\'n}ski, W. (2020). A new exact confidence interval for the difference of two binomial proportions. REV\-STAT-Sta\-tis\-ti\-cal Journal. 18: 521-530

\endgroup

\vfill\eject

{\bf Appendix 1}
\bigskip

A few remarks before the proof.

\bigskip

{\bf Remark 1.} $P_{r}\left\{n_{A1},n_{B1}\right\}\to\cases{0,&as $r\to0$\cr 0,&as $r\to+\infty$\cr}$ for $1\leq n_{A1}\leq n_A-1$ and $1\leq n_{B1}\leq n_B-1$

\medskip
{\it Proof of Remark 1.} For $1\leq n_{A1}\leq n_A-1$ and $1\leq n_{B1}\leq n_B-1$
$$\eqalign{
P_{r,p_A}\left\{n_{A1},n_{B1}\right\}&\propto p_A^{n_{A1}}(1-p_A)^{n_A-n_{A1}}\cdot \left(\frac{p_A}{p_A+r(1-p_A)}\right)^{n_{B1}}\left(\frac{r(1-p_A)}{p_A+r(1-p_A)}\right)^{n_B-n_{B1}}\cr
&\to\cases{
0,&as $r\to0$\cr
0,&as $r\to+\infty$\cr
}
}
$$
Hence $P_{r}\left\{n_{A1},n_{B1}\right\}\to0$ as $r\to0$ or $r\to\infty$.

\bigskip

{\bf Remark 2.} $P_r\{\hor=0\}\to\cases{\frac{n_A}{n_A+1},&as $r\to0$\cr 0,&as $r\to+\infty$\cr}$

\medskip

{\it Proof of Remark 2.} Note that $\hor=0$ iff $(n_{A1}=0$ and $n_{B1}\geq1)$ or $(1\leq n_{A1}\leq n_A-1$ and $n_{B1}=n_B)$. Hence
$$\eqalign{
P_{r,p_A}\{\hor=0\}&=(1-p_A)^{n_A}\sum_{n_{B1}\geq1}\binom{n_B}{n_{B1}}p_B^{n_{B1}}(1-p_B)^{n_B-n_{B1}}+p_B^{n_B}\sum_{n_{A1}=1}^{n_A-1}\binom{n_A}{n_{A1}}p_A^{n_{A1}}(1-p_A)^{n_A-n_{A1}}\cr
&=(1-p_A)^{n_A}\left(1-\left(\frac{r(1-p_A)}{p_A+r(1-p_A)}\right)^{n_B}\right)+\left(\frac{p_A}{p_A+r(1-p_A)}\right)^{n_B}\left(1-p_A^{n_A}-(1-p_A)^{n_A}\right)\cr
&\to\cases{
(1-p_A)^{n_A}+\left(1-p_A^{n_A}-(1-p_A)^{n_A}\right)=1-p_A^{n_A},&as $r\to0$\cr
0,&as $r\to+\infty$\cr
}}$$
We obtain
$$P_r\{\hor=0\}=\int_0^1P_{r,p_A}\{\hor=0\}dp_A\to\cases{
\frac{n_A}{n_A+1},&as $r\to0$\cr
0,&as $r\to+\infty$\cr
}$$

\bigskip

{\bf Remark 3.} $P_r\{\hor=1\}\to\cases{\frac{1}{n_A+1},&as $r\to0$\cr \frac{1}{n_A+1},&as $r\to+\infty$\cr}$
\medskip

{\it Proof of Remark 3.} Note that $\hor=1$ iff $n_{A1}n_B=n_{B1}n_A$. Hence
$$\eqalign{
P_{r,p_A}\{\hor=1\}
&=(1-p_A)^{n_A}(1-p_B)^{n_B}+p_A^{n_A}p_B^{n_B}+\sum_{n_{A1}=1}^{n_A-1}P_{r,p_A}\left\{n_{A1},n_{B1}\right\}\cr
&=(1-p_A)^{n_A}\left(\frac{r(1-p_A)}{p_A+r(1-p_A)}\right)^{n_B}+p_A^{n_A}\left(\frac{p_A}{p_A+r(1-p_A)}\right)^{n_B}+\sum_{n_{A1}=1}^{n_A-1}P_{r,p_A}\left\{n_{A1},n_{B1}\right\}\cr
&\to\cases{
p_A^{n_A},&as $r\to0$\cr
(1-p_A)^{n_A},&as $r\to+\infty$\cr
}}$$
We obtain
$$P_r\{\hor=1\}=\int_0^1P_{r,p_A}\{\hor=1\}dp_A\to\cases{
\frac{1}{n_A+1},&as $r\to0$\cr
\frac{1}{n_A+1},&as $r\to+\infty$\cr
}$$

\bigskip

{\bf Theorem.} For $n_A>\frac{2}{1-\gamma}-1$ the confidence interval for $r$ is two-sided and is one-sided otherwise.

\bigskip

{\it Proof.}

For $0<t<1$ we have
$$P_r\left\{\hor\leq t\right\}=P_r\left\{\hor=0\right\}+P_r\left\{0<\hor\leq t\right\}\to\cases{\frac{n_A}{n_A+1},&as $r\to0$\cr 0,&as $r\to+\infty$\cr}$$
If $\frac{n_A}{n_A+1}>\frac{1+\gamma}{2}$, i.e. $n_A>\frac{2}{1-\gamma}-1$,  the confidence interval is two-sided. Otherwise, the c.i. is one-sided with the left end equal to $0$.

For $1\leq t<+\infty$ we have
$$P_r\left\{\hor\leq t\right\}=P_r\left\{\hor<1\right\}+P_r\left\{\hor=1\right\}+P_r\left\{1<\hor< +\infty\right\}\to\cases{1,&as $r\to0$\cr \frac{1}{n_A+1},&as $r\to+\infty$\cr}$$
If $\frac{1}{n_A+1}<\frac{1-\gamma}{2}$, i.e. $n_A>\frac{2}{1-\gamma}-1$, the confidence interval is two-sided. Otherwise the c.i. is one sided with the right end equal to $+\infty$.

\vfill\eject
{\bf Appendix 2}
\bigskip

An exemplary R code for calculating the confidence interval for the odds ratio is enclosed.
\bigskip

\begingroup
\parindent0pt
\baselineskip10pt
\font\ttt=pltt8
\ttt
\obeylines

OR=function(n,m)$\{$
  ifelse(m[1]==0 \& m[2]==0,0,
    ifelse(m[1]==n[1] \& m[2]==n[2],2*(n[1]-1)*(n[2]-1),
      ifelse(m[2]==0,2*(n[1]-1)*(n[2]-1),
        ifelse(m[1]==n[1],2*(n[1]-1)*(n[2]-1),m[1]*(n[2]-m[2])/(n[1]-m[1])/m[2])
        )))$\}$
\smallskip

f=function(rr,k1,k2,pA)$\{$dbinom(k1,n[1],pA)*dbinom(k2,n[2],pA/(pA+rr*(1-pA)))$\}$
\smallskip

nieostra=function(rr,tt)$\{$
  line<-0
  prawd=c()
  for (k1 in 0:(n[1]-1))$\{$
    RS=round(n[2]/(tt*(n[1]/k1-1)+1),2)
    Niod=ifelse(k1==0,ifelse(tt<1,1,0),ceiling(RS))
    for (k2 in Niod:n[2])
    $\{$mrob=c(k1,k2)
    line=line+1;
    prawd[line]=integrate(f,0,1,rr=rr,k1=k1,k2=k2,subdivisions = 1000L,stop.on.error = FALSE)\$value;$\}\}$
    td=sum(prawd)$\}$

\smallskip

ostra=function(rr,tt)$\{$
  line<-0
  prawd=c()
  for (k1 in 0:(n[1]-1))$\{$
    RS=round(n[2]/(tt*(n[1]/k1-1)+1),2)
    Osod=ifelse(k1==0,ifelse(tt<=1,1,0),ifelse(RS==trunc(RS),RS+1,ceiling(RS)))
    for (k2 in Osod:n[2])
    $\{$mrob=c(k1,k2)
    line=line+1;
    prawd[line]=integrate(f,0,1,rr=rr,k1=k1,k2=k2,subdivisions = 1000L,stop.on.error = FALSE)\$value;$\}\}$
  tg=sum(prawd)$\}$

\smallskip

CI=function(n,m,level)$\{$
  orobs<-OR(n,m)
  eps=1e-4

  ifelse(orobs<1,
         $\{$ifelse(n[1]<=2/(1-level)-1,
           $\{$L=0;
           P=uniroot(function(t)$\{$ostra(t,orobs)-(1-level)/2$\}$, lower = orobs, upper = 2*(n[1]-1)*(n[2]-1),
           \hskip1em tol = eps)\$root$\}$,
           $\{$L=uniroot(function(t)$\{$nieostra(t,orobs)-(1+level)/2$\}$, lower = 0.00000001, upper = orobs,
           \hskip1em tol = eps)\$root;
           P=uniroot(function(t)$\{$ostra(t,orobs)-(1-level)/2$\}$, lower = orobs, upper = 2*(n[1]-1)*(n[2]-1),
           \hskip1em tol = eps)\$root$\}$)$\}$,
         $\{$ifelse(n[1]<=2/(1-level)-1,
                 $\{$L=uniroot(function(t)$\{$nieostra(t,orobs)-(1+level)/2$\}$, lower = 0.00000001, upper = orobs, tol = eps)\$root;
                 P=Inf$\}$,
                 $\{$L=uniroot(function(t)$\{$nieostra(t,orobs)-(1+level)/2$\}$, lower = 0.00000001, upper = orobs, tol = eps)\$root;
                 P=uniroot(function(t)$\{$ostra(t,orobs)-(1-level)/2$\}$, lower = orobs, upper = 2*(n[1]-1)*(n[2]-1),
          \hskip1em tol = eps)\$root$\}$)$\}$
         )
    print(paste("Confidence interval for odds ratio (",round(L,5),",",round(P,5),") at the confidence level ",
    \hskip1em level,sep=""),quote=FALSE)
    print(paste("Sample odds ratio equals ",round(orobs,4), "; n1=",n[1],", n2=",n[2],sep=""),quote=FALSE)
$\}$
\smallskip

\#Example of usage
n=c(60,70) \# input $\scriptstyle n_A$ and $\scriptstyle n_B$
m=c(7,63) \# input $\scriptstyle n_{A1}$ and $\scriptstyle n_{B1}$
CI(n,m,level=0.95)

\endgroup

\bye